\documentclass[conference]{IEEEtran}
\IEEEoverridecommandlockouts

\usepackage{cite}
\usepackage{amsmath,amssymb,amsfonts}
\usepackage{graphicx}
\usepackage{textcomp}
\usepackage{xcolor}
\def\BibTeX{{\rm B\kern-.05em{\sc i\kern-.025em b}\kern-.08em
    T\kern-.1667em\lower.7ex\hbox{E}\kern-.125emX}}
\makeatletter
\newcommand{\linebreakand}{%
  \end{@IEEEauthorhalign}
  \hfill\mbox{}\par
  \mbox{}\hfill\begin{@IEEEauthorhalign}
}    
\usepackage{graphicx}
\usepackage{xcolor}

\usepackage{mathtools}
\usepackage{cuted}
\usepackage{eqnarray, amsmath}
\usepackage{amssymb}
\usepackage{multicol,lipsum}
\usepackage{cite}
\usepackage[colorlinks=Flase]{hyperref}
\usepackage{bbm}
\usepackage{algorithm} 
\usepackage{algpseudocode}
\usepackage{caption}
\usepackage{steinmetz}
\usepackage{subcaption}
\usepackage{amsthm}
\usepackage[english]{babel}

\newcounter{relctr} 
\everydisplay\expandafter{\the\everydisplay\setcounter{relctr}{0}} 
\DeclareMathOperator*{\maximize}{maximize} 
\usepackage[figurename=Fig.]{caption}

\begin{document}

\title{On the Radio Stripe Deployment for Indoor RF Wireless Power Transfer\\
\thanks{This work is partially supported in Finland by the Finnish Foundation for Technology Promotion, the Research Council of Finland (former Academy of Finland) 6G Flagship Programme (Grants 348515 and 346208 (6G Flagship)) and by the European Commission through the Horizon Europe/JU SNS project Hexa-X-II (Grant Agreement no. 101095759)}
}

\author{\IEEEauthorblockN{1\textsuperscript{st} Amirhossein~Azarbahram}
\IEEEauthorblockA{\textit{Centre for Wireless Communications} \\
\textit{University of Oulu}, 
Oulu, Finland \\
amirhossein.azarbahram@oulu.fi}
\and
\IEEEauthorblockN{2\textsuperscript{nd} Onel~L.~A.~López}
\IEEEauthorblockA{\textit{Centre for Wireless Communications} \\
\textit{University of Oulu}, 
Oulu, Finland \\
onel.alcarazlopez@oulu.fi}
\linebreakand 
\and
\IEEEauthorblockN{3\textsuperscript{rd} Petar~Popovski}
\IEEEauthorblockA{\textit{Department of Electronic Systems} \\
\textit{Aalborg University}, 
Aalborg, Denmark \\
petarp@es.aau.dk}
\and
\IEEEauthorblockN{4\textsuperscript{th} Matti~Latva-Aho}
\IEEEauthorblockA{\textit{Centre for Wireless Communications} \\
\textit{University of Oulu}, 
Oulu, Finland \\
matti.latva-aho@oulu.fi}
}

\maketitle

\begin{abstract}
One of the primary goals of future wireless systems is to foster sustainability, for which, radio frequency (RF) wireless power transfer (WPT) is considered a key technology enabler. The key challenge of RF-WPT systems is the extremely low end-to-end efficiency, mainly due to the losses introduced by the wireless channel. Distributed antenna systems are undoubtedly appealing as they can significantly shorten the charging distances, thus, reducing channel losses. Interestingly, radio stripe systems provide a cost-efficient and scalable way to deploy a distributed multi-antenna system, and thus have received a lot of attention recently. Herein, we consider an RF-WPT system with a transmit radio stripe network to charge multiple indoor energy hotspots, i.e., spatial regions where the energy harvesting devices are expected to be located, including near-field locations. We formulate the optimal radio stripe deployment problem aimed to maximize the minimum power received by the users and explore two specific predefined shapes, namely the straight line and polygon-shaped configurations. Then, we provide efficient solutions relying on geometric programming to optimize the location of the radio stripe elements. The results demonstrate that the proposed radio stripe deployments outperform a central fully-digital square array with the same number of elements and utilizing larger radio stripe lengths can enhance the performance, while increasing the system frequency may degrade it.
\end{abstract}


\vspace{-2mm}
\section{Introduction}
\vspace{-1mm}
\IEEEPARstart{R}{adio} frequency (RF) wireless power transfer (WPT) is a promising technology for future wireless systems. Indeed, RF-WPT may enable uninterrupted communication by preventing battery depletion in the devices. Notably, RF-WPT systems can provide wireless charging capability over large distances while utilizing the same infrastructure as wireless communication \cite{intro1,intro2}. However, the main drawback of these systems is the low end-to-end power transfer efficiency, mainly due to the huge losses introduced by the wireless channel. This has motivated the exploitation of novel techniques relying, e.g., on energy beamforming and distributed antenna systems \cite{intro3}.

Beamforming may be performed in a digital, analog, or hybrid way depending on the transmitter architecture. Although digital beamforming requires a relatively large number of RF chains, which leads to high-cost implementations, it provides the highest degrees of freedom in terms of focusing the beams toward the desired directions and compensating for the signal propagation loss \cite{hybridbeamsurvey}. On the other hand, while analog beamforming reduces the number of RF chains by utilizing analog circuits, it sacrifices flexibility for cost reduction. Meanwhile, hybrid architectures are introduced to provide a trade-off between cost and flexibility by combining both aforementioned architectures. Notably, conventional multi-antenna systems with co-located antenna elements may still suffer from significant path loss, even after leveraging beamforming. This comes from the close proximity of all the elements in the array, resulting in nearly identical paths for the transmitted signals. Consequently, if one of these paths is obstructed or subjected to a long distance to the user, all other elements in the array are also affected in a similar manner, unless the array size is extremely large. To cover the areas suffering from non-line-of-sight conditions, intelligent reflecting surfaces with passive reflective elements could be deployed \cite{IRS-basis}. However, the signal attenuation between the transmitter and the reflecting elements may considerably limit the potential performance gains. 

Instead, distributively deploying transmit antenna elements across the area for RF-WPT may more actively reduce blind spots and shorten the charging distances. For this, it is critical to optimize the position of the antenna elements based on the characteristics of the deployment area and user density in different locations. For instance, the authors in \cite{distributedWPT1} propose a method to find the radius of a circle, on which antenna power beacons (PB) are uniformly distributed, aiming to maximize the efficiency of an RF-WPT system. Meanwhile, the impact of the number of PBs on the minimum received energy is investigated in \cite{intro3}, where the locations of PBs are optimized using the K-means clustering algorithm. In \cite{Osmel_Deployment}, multiple approaches are proposed to optimize the position of PBs aiming to maximize the average energy received in the worst location of the area.

Interestingly, the radio stripe system provides a cost-efficient way to distribute the antenna elements in the area. In this setup, the antenna elements and their corresponding processing units are placed along a cable. Therefore, a long enough radio stripe may equip many antenna elements deployed throughout the area, which can reduce the expected distance to the users and avoid blockage \cite{intro_radiostripe}. Another benefit of the radio stripe system is its ability to provide a large diameter, i.e., the largest size of the antenna aperture. Therefore, radio stripe systems can more easily create near-field conditions than conventional transmitter architectures in wireless communications. Notice that when operating in the near-field region, the wavefronts incident upon a receiving node may exhibit a strictly spherical nature, providing the capability to focus the RF power on specific spatial points rather than just spatial directions, which may result in more efficient energy delivery \cite{near-field}.

In addition, future RF-WPT applications may mostly occur indoors with predefined hotspots where the power demand is expected to be high, for which radio stripe setups can be appealing. Imagine, for instance, a restaurant as in Fig.~\ref{fig:system_model}, where wearables and laptops are usually located close to the tables and the counter, thus, these are potential hotspots. Meanwhile, it is crucial to properly design the radio stripe network and choose the location of the elements to further promote scalability and mitigate channel loss. Although there are some works on signal processing for radio stripe systems \cite{OnelRadioStripes, shaik2021mmseRadiostripe}, to the best of our knowledge, no work has yet attempted to optimize the location of the radio stripe antenna elements. Herein, we aim precisely to fill this research gap. 

Our main contributions are three-fold: i) we formulate the radio stripe deployment problem for an RF-WPT system transmitting energy signals to charge multiple hotspot locations, where devices may often request charging services, while considering that these locations could be in the near-field radiative region of the radio stripe; ii) we consider polygon-shaped and line-shaped radio stripe networks to reduce the problem complexity, and propose an efficient framework relying on geometric programming (GP) to derive the locations of the antenna elements and; iii) we show numerically that the proposed deployments can outperform a central fully-digital square array in terms of minimum power received by the users and that increasing the radio stripe length is beneficial since more degrees of freedom are available for beamforming and for distributing the elements across the area.


The remainder of the paper is structured as follows. Section~\ref{sys and prob} introduces the system model and the problem formulation. The proposed optimization framework is discussed in Section~\ref{sec:Optimization}, while Section~\ref{result} provides the numerical analysis and Section~\ref{conclusion} concludes the paper.


\section{System Model and Problem Formulation}\label{sys and prob}

\begin{figure}
    \centering
    \includegraphics[width=0.9\columnwidth]{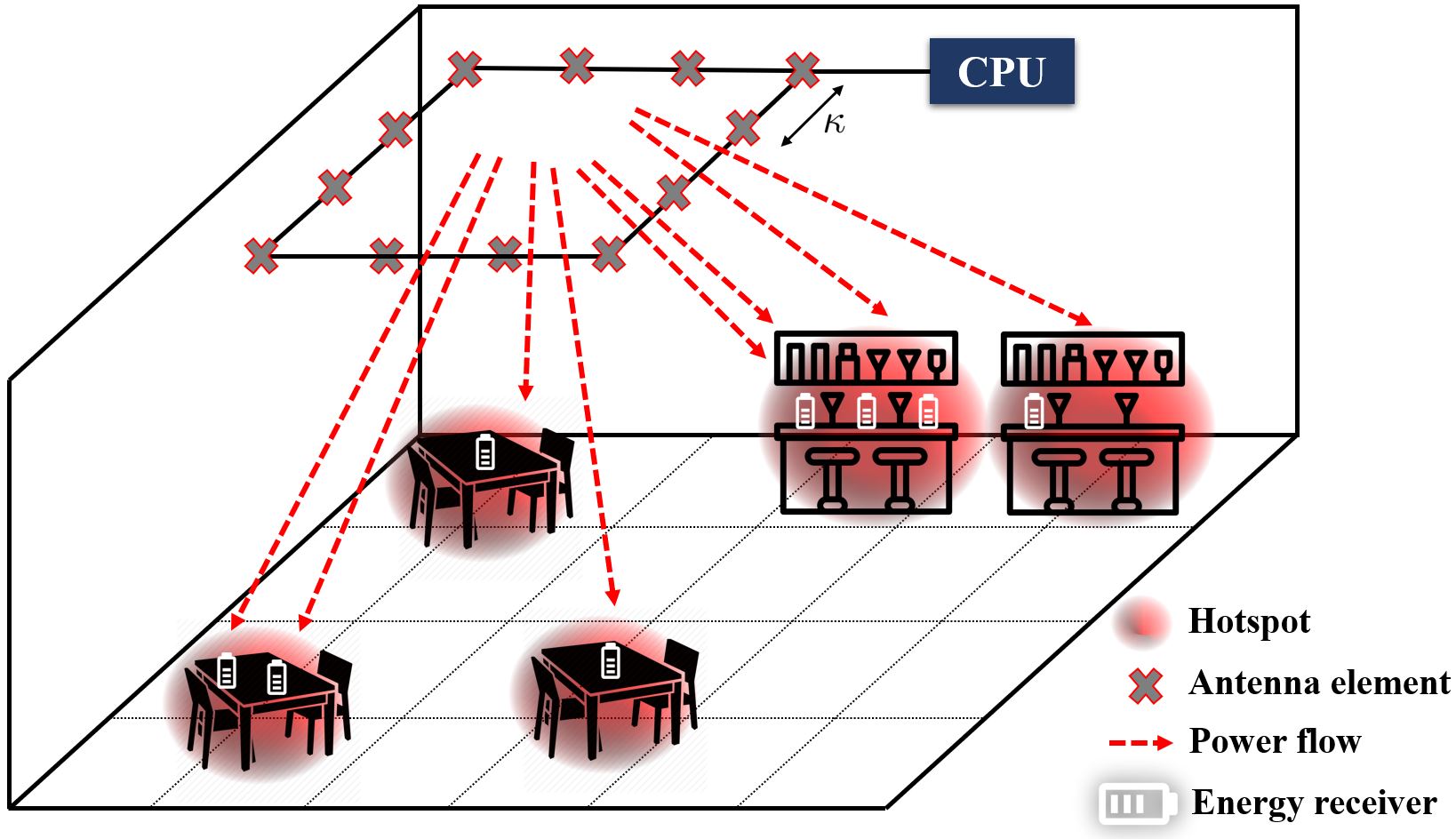}
    \vspace{-2mm}
    \caption{Radio stripe system model with a central processing unit (CPU), exemplified with a restaurant scenario.}
    \label{fig:system_model}
    \vspace{-5mm}
\end{figure}

We consider a multi-antenna RF-WPT system where the transmitter is deployed using a fully-digital radio stripe network and the energy harvesting devices are clustered around $M$ hotspots. The radio stripe network consists of $N \geq M$ elements with an inter-element distance $\kappa$ and located on the ceiling with height $h_c$. Moreover, $\mathbf{g}_j = \begin{bmatrix} g_{j,1}, g_{j, 2}, g_{j, 3} \end{bmatrix}^T$ with $g_{j, 3} = h_c$ is the Cartesian coordinate of the $j$th element, and $\mathbf{q}_i = \begin{bmatrix} q_{i,1}, q_{i,2}, q_{i,3} \end{bmatrix}^T$ denotes the location of the center of $i$th hotspot. Note that in the rest of this paper, the term hotspot refers to the center of a hotspot. The system model is depicted in Fig.~\ref{fig:system_model} by adopting a conventional restaurant configuration, featuring designated hotspot areas where the devices are expected to be positioned. 

\subsection{Channel Model}

We consider RF-WPT within indoor environments, thus we adopt a near-field line-of-sight wireless channel model \cite{near-field}. Let $D$ be the antenna diameter, then, user $i$ at a distance $r_i$ from the transmitter lies in the radiative near-field region if
\begin{equation}\label{eq:nearfield}
    \sqrt[3]{D^4/(8\lambda)} = r_{fs} < r_i < r_{fr} = 2D^2/\lambda,
\end{equation}
where $\lambda = \frac{C}{f}$ is the wavelength at frequency $f$ with $C$ being the speed of light. Additionally, $r_{fs}$ and $r_{fr}$ are the Fresnel and Fraunhofer distance, respectively. Thus, it is evident that the near-field region can be expanded by increasing both the system frequency and the size of the antenna array.

The channel coefficient between element $j$ and hotspot $i$ is given by
\begin{equation}\label{eq:channelcoef}
    \gamma_{j, i} = A_{j, i} e^{\frac{-j2\pi}{\lambda} ||\mathbf{g}_j - \mathbf{q}_i||},
\end{equation}
where the term $2\pi ||\mathbf{g}_j - \mathbf{q}_i||/\lambda$ represents the phase shift that is introduced due to the propagation distance and $||\mathbf{g}_j - \mathbf{q}_i||$ is the channel length. Moreover, $A_{j, i}$ is the corresponding channel gain, which is given by
\begin{equation}
    A_{j, i} = \sqrt{F(\theta_{j, i})}\frac{\lambda}{4\pi ||\mathbf{g}_j - \mathbf{q}_i||}.
\end{equation}
Herein, $F(\theta_{j, i})$ is the antenna radiation profile given by \cite{anetnna_radiation}
\begin{equation}
        F(\theta_{j, i}) = \begin{cases}
        2(b + 1){\cos^b {\theta_{j, i}}}, & \theta_{j, i} \in [0,\pi/2],
        \\
        0, & \text{otherwise},
        \end{cases}
\end{equation}
where $b$ is the the boresight gain and $\theta_{j, i}$ is the elevation angle between the $j$th element and the $i$th hotspot. Since the antenna elements are located at the ceiling, we can write 
\begin{equation}\label{eq:cos}
    \cos{\theta_{j,i}} = (h_c - q_{i, 3})/(||\mathbf{g}_j - \mathbf{q}_i||),\quad \theta_{j, i} \in [0, \pi/2].
\end{equation}
Also, $\boldsymbol{\gamma}_i \!=\! \begin{bmatrix}
\gamma_{1,i} ,\gamma_{2,i}, \ldots, \gamma_{N,i}\end{bmatrix}^T \!\in\! \mathbb{C}^{N \times 1}$ collects the channel coefficients between hotspot $i$ and the antenna elements. 

Note that the near-field channel model is also capable of capturing the far-field condition. Specifically, when the hotspot is in the far-field region, the channel coefficient is simplified to $A_{i} e^{-j\psi_{j,i}}$, where $A_i$ is contingent solely on the distance between the $i$th hotspot and the transmitter, while $\psi_{j, i}$ is exclusively determined by the user's direction and the spatial configuration of the antenna elements within the array.

Let us consider $M$ independent and normalized energy symbol, while the energy beams are focused toward the hotspots using digital beamforming \cite{OnelRadioStripes}. Herein, $\mathbf{w}_m \in \mathbb{C}^{N\times 1}$ is the digital precoder corresponding to the $m$th energy symbol, while the power of the RF signal received by the $i$th hotspot and averaged out over the signal waveform is given by
\begin{equation}\label{eq:rxpower}
    P^{rx}_{i} = \sum\nolimits_{m = 1}^{M} ||\boldsymbol{\gamma}_i^H \mathbf{w}_m||^2.
\end{equation}
\vspace{-6mm}
\subsection{Problem Formulation}
\vspace{-1mm}
The goal is to simultaneously deploy the radio stripe system and design the beamforming such that the hotspots are served fairly. To accomplish this, the minimum received RF power by the hotspots has to be maximized, and thus, the optimization problem can be formulated as
\begin{subequations}\label{prob_basic}
\begin{align}
\label{prob_basic_a}  \quad \maximize_{\{\mathbf{g}_j\}_{\forall j}, \{\mathbf{w}_i\}_{\forall i}} & \min_{i} P^{rx}_{i}/k_i \\ 
\textrm{subject to} \label{prob_basic_b} \quad & \sum\nolimits_{i = 1}^{M} ||\mathbf{w}_i||^2 \leq \Tilde{P}, \\
& \label{prob_basic_c} \sum\nolimits_{j = 1}^{N - 1} ||\mathbf{g}_j - \mathbf{g}_{j+1}|| \leq (N - 1)\kappa, \\
&  \label{prob_basic_d} ||\mathbf{g}_j - \mathbf{g}_n|| \geq \kappa, \quad \forall j,n\ \text{with}\ {j\neq n},
\end{align}
\end{subequations}
where $\Tilde{P}$ is the transmit power budget and $k_i$ is the user density, i.e., the expected number of devices, in the $i$th hotspot. Furthermore, the objective \eqref{prob_basic_a} is non-convex, and \eqref{prob_basic_b} is a convex constraint imposing that the maximum input power should not exceed $\Tilde{P}$. Additionally, the constraints \eqref{prob_basic_c} and \eqref{prob_basic_d} impose practical limitations on the radio stripe, requiring consecutive elements within the array to be equidistant from each other, and ensuring that the minimum distance between any two elements is at least $\kappa$. Herein, \eqref{prob_basic_c} is a convex constraint, while \eqref{prob_basic_d} is non-convex. 

Note that considering beamforming optimization in this problem is only for deployment purposes. In practice, the devices in the hotspots may be served by dedicated beams given a radio stripe deployment as the number of antenna elements is typically greater than the number of devices. Mathematically, this implies that considering $R$ users located in the hotspots, we have $ P^{rx}_{i} = \sum_{m=1}^R ||\boldsymbol{\gamma}_i^H \mathbf{w}_m||^2$ as the received power by the $i$th user.
\vspace{-2mm}
\section{Optimization Framework} \label{sec:Optimization}
\vspace{-1mm}
It is known that for a given radio stripes deployment, the optimal precoders can be derived by transforming \eqref{prob_basic} into a semi-definite program (SDP) \cite{onellowcomp}. As a result, each precoder depends heavily on all the channel coefficients, thus, making the simultaneous optimization of antenna locations and precoders highly complex. To cope with this, we choose maximum ratio transmission (MRT)-based precoders, which can reduce the complexity of \eqref{prob_basic} due to its well-defined structure. Notice that the MRT-based precoder serves as a local optimum solution for the beamforming design problem \cite{OnelRadioStripes}, and thus, provides a lower bound for the optimization objective. Specifically, $\mathbf{w}_m^\star = \frac{\boldsymbol{\gamma_{m}}}{|\boldsymbol{\gamma_{m}}|}\sqrt{P_m}$ is the $m$th MRT-based precoder, where $P_m$ is the assigned power to the $m$th precoder. Hereby, \eqref{eq:rxpower} can be reformulated as

\begin{equation} \label{eq:rxpower2}
    P^{rx}_{i} = \sum_{m = 1}^{M} \bigl|\bigl|\boldsymbol{\gamma}_i^H \frac{\boldsymbol{\gamma}_m}{|\boldsymbol{\gamma}_m|}\sqrt{P_m}\bigr|\bigr|^2.
\end{equation}
Even considering this, it is difficult to solve the problem because of its highly non-linear objective function. Specifically, the term $e^{\frac{-j2\pi}{\lambda} ||\mathbf{g}_j - \mathbf{q}_i||}$ is an oscillating function of the distance. As previously mentioned, the devices are mostly served by dedicated beams in practical beamforming designs. Thus, we discard the impact of non-dedicated beams to lower bound \eqref{eq:rxpower2} as $P_i^{rx}> P_i ||\gamma_i||^2$ and utilize it to further reduce the complexity of the problem, which can be rewritten as
\begin{subequations}\label{prob_MRT}
\begin{align}
\label{prob_MRT_a}  \quad \maximize_{\{\mathbf{g}_j\}_{\forall j}, \{P_i\}_{\forall i}} \quad & \min_{i} {P_i} ||\boldsymbol{\gamma}_i||^2/k_i \\ 
\textrm{subject to} \label{prob_MRT_b} \quad & \sum\nolimits_{i = 1}^{M} P_i \leq \Tilde{P}, \\
&  \eqref{prob_basic_c}, \eqref{prob_basic_d}. \nonumber 
\end{align}
\end{subequations}
Next, by utilizing \eqref{eq:channelcoef}-\eqref{eq:cos}, we can write 
\begin{align}\label{eq:gain_simp}
    ||\boldsymbol{\gamma}_i||^2 &= \sum\nolimits_{j = 1}^{N} \Bigl(\sqrt{2(b + 1){\cos^b {\theta_{j,i}}}}\frac{\lambda}{4\pi ||\mathbf{g}_j - \mathbf{q}_i||}\Bigr)^2 \nonumber \\
    &= 2(b \!+\! 1)(h_c\!-\!q_{i, 3})^b\biggl(\frac{\lambda}{4\pi}\biggr)^2 \sum_{j = 1}^{N} \frac{1}{||\mathbf{g}_j\!-\!\mathbf{q}_i||^{b\!+\!2}}.
\end{align}
Since the terms $\frac{\lambda}{4 \pi}$ and $2(b+1)$ have no impact on the optimization, problem \eqref{prob_MRT} can be reformulated as 
\begin{subequations}\label{prob_MRT_final}
\begin{align}
\label{prob_MRT_final_a} \quad \maximize_{\substack{t, \{\mathbf{g}_j\}_{\forall j} \\ \{P_i\}_{\forall i}}} \quad & t \\[-4pt]
\textrm{subject to} \quad & \label{prob_MRT_final_b}  t   \leq \sum\nolimits_{j = 1}^N \frac{e_{i}^{b} P_i}{k_i ||\mathbf{g}_j - \mathbf{q}_i||^{b+2}}, \quad \forall i\\
& \eqref{prob_basic_c}, \eqref{prob_basic_d}, \nonumber\eqref{prob_MRT_b},
\end{align}
\end{subequations}
where $e_{i} = h_c - q_{i, 3}$.

One way to reduce the complexity of the deployment problem is to consider predetermined shapes, thus pre-addressing constraints \eqref{prob_basic_c} and \eqref{prob_basic_d}. Notably, using predefined shapes may be more reasonable for practical implementations as it simplifies the manufacturing and deployment of radio stripes, rather than dealing with distorted shapes.

\begin{figure}[t]
\centering
    \centering
    \includegraphics[width=\columnwidth]{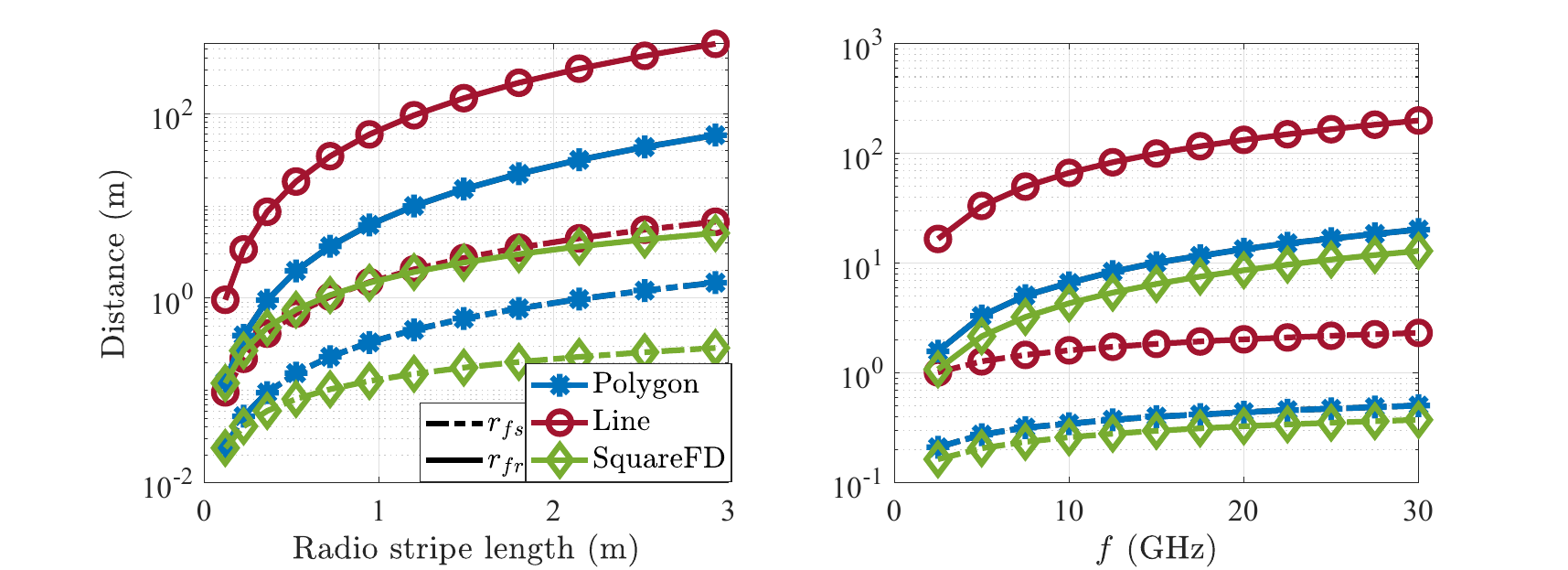}
    \caption{Fraunhofer and Fresnel distances as a function of (a) the radio stripe length with $f = 10$ GHz (left) and (b) the frequency for a 1 m radio stripe length (right). The Square-FD refers to a square planar array with its number of elements matching the nearest square number to the number of radio stripe elements for that given length/frequency.}
    \label{nearfield}
    \vspace{-5mm}
\end{figure}

Herein, we consider two shapes for radio stripe deployment: regular polygon and straight line. It is evident that for a regular polygon with $N$ elements positioned at the edges and denoting $r_0 = \kappa/({2 \sin{\frac{\pi}{N}}})$ as the distance between the center and elements, one has $D \le 2 r_0 =\kappa/\sin{\frac{\pi}{N}}$, which becomes tight as $N$ increases while converging to $\kappa N/\pi$. On the other hand, a straight line provides the largest antenna diameter, which is the length of the line, i.e., $(N - 1)\kappa$. Therefore, a straight line-shaped radio stripe makes it considerably easy to create near-field conditions. Fig.~\ref{nearfield} visually corroborates the discussion about near-field region for different transmit architectures. Meanwhile, one drawback of the line-shaped deployment is that some elements might have to be positioned at large distances from specific users leading to limitations for beam focusing, while this is not the issue in the polygon case. Thus, there exists a trade-off between creating near-field conditions and beamforming performance depending on the area and the user distribution.

%
\vspace{-2mm}
\subsection{Ploygon-shaped Radio Stripe}
\vspace{-1mm}
 Herein, we consider the radio stripe to be shaped like a regular polygon. Assuming this, all of the elements can be positioned according to their relation with the center's location and the rotation angle. Although the location of the elements changes with rotation, the resolution of such rotation is upper bounded by $\phi = {2\pi}/{N}$, thus, it becomes negligible for a large $N$, as typical in radio stripe systems. Hence, we discard the influence of the rotation angle and proceed by defining $\hat{\mathbf{g}} = \begin{bmatrix}
    \hat{g}_1, \hat{g}_2, \hat{g}_3
\end{bmatrix}^T$ with $\hat{g}_3 = h_c$ as the location of the center of a regular polygon, which is placed on the ceiling. Then, we can write the location of the $j$th element as
\begin{align}\label{polyeqq}
      \big\{g_{j, 1} \!=\! \hat{g}_1\!+\!r_0\cos{(j\!-\!1)\phi},\ g_{j, 2}\!=\!\hat{g}_2\!+\!r_0\sin{(j\!-\!1)\phi}\big\}.
\end{align}
 Let us proceed by defining $\hat{\mathbf{q}}_{j, i} = \mathbf{q}_i - \begin{bmatrix}
    r_0\cos{(j - 1)\phi}, r_0\sin{(j - 1)\phi}, 0
\end{bmatrix}^T$ and $\hat{d}_{j, i} = ||\hat{\mathbf{g}} - \hat{\mathbf{q}}_{j, i}||$. Hereby, the optimization problem can be formulated as 
\begin{subequations}\label{prob_sgp_ploy}
\begin{align}
\label{prob_sgp_ploy_a} \quad \maximize_{\substack{\{P_i, \hat{d}_{j,i}\}_{\forall j, i}\\ t, \hat{g}_{1}, \hat{g}_{2}}} \quad & t \\[-6pt] 
\textrm{subject to} \quad & \label{prob_sgp_ploy_b} k_i e_{i}^{-b} P_i^{-1} t   \leq \sum\nolimits_{j = 1}^N \hat{d}_{j,i}^{-(b+2)}, \quad \forall i, \\
\label{prob_sgp_ploy_d} \quad & \hat{d}_{j,i}^{-2} \bigl(\sum_{u} {\hat{g}_{u}}^2 +  \sum_{u} {\hat{q}_{j,i,u}}^2\bigr) \nonumber \\ 
    & \leq 1 + 2\hat{d}_{j,i}^{-2} \sum_{u} \hat{g}_{u}\hat{q}_{j,i,u}, \quad \forall j,i, \\[-6pt]
    &    \eqref{prob_MRT_b}, \nonumber 
\end{align}
\end{subequations}
where $\hat{q}_{j,i,u}$ is the $u$th element in $\hat{\mathbf{q}}_{j, i}$ and \eqref{prob_sgp_ploy_d} comes from squaring both sides and expanding the inequality $\hat{d}_{j,i} \geq ||\mathbf{\hat{g}} - \hat{\mathbf{q}}_{j, i}||$. Note that minimizing $\hat{d}_{j,i}\ \text{with}\ \hat{d}_{j,i}\ge 0, \forall j,i,$ will lead to maximizing $t$. Thus,  \eqref{prob_sgp_ploy_d} forces $\hat{d}_{j,i}$ to be equal to the distance between hotspot $i$ and element $j$.

Notice that \eqref{prob_sgp_ploy} is a signomial programming (SGP) problem, and thus can be efficiently solved by relaxing it as a GP problem, although without global optimality guarantees \cite{gp_boyd}. Herein, \eqref{prob_sgp_ploy} can be transformed into a standard GP by utilizing local monomial approximation near the point $\hat{g}_{1}^{(0)}, {\hat{g}_{2}}^{(0)}, \{\hat{d}_{j,i}^{(0)}\}_{\forall j,i}$ as 
\begin{subequations}\label{prob_gp_ploy}
\begin{align}
\label{prob_gp_ploy_a}  \quad \maximize_{\substack{\{P_i, \hat{d}_{j,i}\}_{\forall j, i}\\ t, \hat{g}_{1}, \hat{g}_{2}}} \quad & t \\[-4pt] 
\textrm{subject to} \quad & \label{prob_gp_ploy_b} k_i e_{i}^{-b} P_i^{-1} t   \leq \hat{h}_{i} \quad \forall i, \\
\label{prob_gp_ploy_d} \quad & \hat{d}_{j,i}^{-2} \bigl(\sum_{u} {\hat{g}_{u}}^2 +  \sum_{u} {\hat{q}_{j,i,u}}^2\bigr) \leq \Tilde{h}_{j, i},  \forall j,i, \\
& \label{prob_gp_ploy_e} \hat{d}^{(0)}_{j,i}/\omega \leq \hat{d}_{j,i} \leq \omega \hat{d}^{(0)}_{j,i}, \quad \forall j,i, \\
& \label{prob_gp_ploy_f} {\hat{g}}^{(0)}_{u}/\omega \leq {\hat{g}}_{u} \leq \omega {\hat{g}}^{(0)}_{ u}, \quad \forall u, \\
&    \eqref{prob_MRT_b}, \nonumber 
\end{align}
\end{subequations}
where $\omega > 1$ sets the approximation trust region \cite{gp_boyd} and 
\begin{align}
&\hat{h}_{i}\! =\! \sum_{j = 1}^N{(\hat{d}^{(0)}_{j,i}})^{-(b+2)} \prod_{l = 1}^N \biggl(\frac{\hat{d}_{l,i}}{\hat{d}^{(0)}_{l,i}} \biggr)^{\hat{\beta}_{l, i}}, \label{eq:hhat}\\
&\hat{\beta}_{l, i}\! =\!  -(b+2)({\hat{d}^{(0)}_{l,i}})^{-(b+2)}\Big/\sum\nolimits_{j = 1}^N{(\hat{d}^{(0)}_{j,i}})^{-(b+2)}, \label{eq:betahhat}\\
& \label{eq:htild1} \Tilde{h}_{j, i}\!=\! \biggl(\!1 \!+\! 2{(\hat{d}^{(0)}_{j,i}})^{-2} \sum_{u} \hat{g}^{(0)}_{u}\hat{q}_{j,i,u}\!\biggr)\biggl(\frac{\hat{d}_{j,i}}{{\hat{d}^{(0)}}_{j,i}}\biggr)^{\Tilde{\beta}_{j, i}}  \prod_{l = 1}^{2} \biggl(\frac{\hat{g}_{l}}{\hat{g}^{(0)}_{l}}\biggr)^{\bar{\beta}_{j,i,l}}\!\!\!\!\!\!, \\
& \label{eq:htild2}\Tilde{\beta}_{j, i}\! =\!-\!4({{\hat{d}^{(0)}}_{l,i}})^{-2} \!\sum_{u} \hat{g}^{(0)}_{u}\hat{q}_{j,i,u}\Big/\!\Big(\!1\!+\!2({{\hat{d}}^{(0)}_{j,i}})^{-2} \sum_{u} \hat{g}^{(0)}_{u}\hat{q}_{j,i,u}\!\Big), \\
& \label{eq:htild3} \bar{\beta}_{j, i, l}\! =\!2\hat{g}^{(0)}_{l}({{\hat{d}^{(0)}}_{j,i}})^{-2} \hat{q}_{j,i,l}\Big/\!\Big(\!1\!+\!2({{\hat{d}^{(0)}}_{j,i}}\!)^{-2}\! \sum_{u}\! \hat{g}^{(0)}_{u}\hat{q}_{j,i,u}\!\Big) .
\end{align}
Notice that \eqref{prob_gp_ploy} is a GP problem, which can be solved efficiently by standard convex optimization tools, e.g., CVX \cite{cvxref}. Algorithm~\ref{sgp_poly_alg} illustrates the proposed optimization approach for polygon-shaped radio stripe deployment. First, a location for the center is chosen, e.g., the center of the area, and $\{\hat{d}_{j,i}^{(0)}\}_{\forall j,i}$ initialized accordingly. Then, the solution and its neighborhood are iteratively updated until the number of iterations reaches $I_{max}$ or the change in the objective function becomes smaller than a specified threshold, $\epsilon$.

GP problems can be solved efficiently in polynomial time using primal-dual interior-point methods \cite{gp_boyd, boyd2004convex}. Moreover, the proposed solutions consist of $I_{max}$ iterations of solving GP problems in the worst case. Therefore, both proposed solutions can be solved efficiently in polynomial time.

\begin{algorithm}[t]
	\caption{Polygon-shaped radio stripe deployment.} \label{sgp_poly_alg}
	\begin{algorithmic}[1]
            \State \textbf{Input:} $\hat{g}_{1}^{(0)}, {\hat{g}_{2}}^{(0)}$, $I_{max}$, $\epsilon$ 
            \State \textbf{Output:} $\{\mathbf{g}_{j}\}_{\forall j}$, $\{P_i\}_{\forall i}$
            \State \textbf{Initialize:} Compute $\{\hat{d}_{j,i}^{(0)}\}_{\forall j,i}$  for $\hat{g}_{1}^{(0)}, {\hat{g}_{2}}^{(0)}$, $iter = 1$, $t =\infty $
            \Repeat
                \State $t' \leftarrow t$ 
                \State Solve \eqref{prob_gp_ploy} to obtain $\hat{g}_{1}, {\hat{g}_{2}}$,  $\{d_{j,i}\}_{\forall j, i}$, and $t$
                \State $\hat{g}^{(0)}_{1} \leftarrow \hat{g}_{1}$, $\hat{g}^{(0)}_{2} \leftarrow \hat{g}_{2}$, $\hat{d}^{(0)}_{j,i} \leftarrow \hat{d}_{j,i}, \quad \forall j, i$
                \State $iter \leftarrow iter + 1$
            \Until{$||t - t'|| \leq \epsilon$ or $iter = I_{max}$}
            \State Obtain $\{\mathbf{g}_{j}\}_{\forall j}$ for $\{\hat{g}_1, \hat{g}_2\}$ using \eqref{polyeqq}
\end{algorithmic} 
\end{algorithm}

\vspace{-2mm}
\subsection{Line-shaped Radio Stripe}
\vspace{-1mm}
Herein, we consider a straight-line-based radio stripe deployment, which is probably the most straightforward implementation. Notably, the corresponding deployment optimization problem is more complex than the polygon-shaped problem. The reason is that the horizontal angle of the line plays a crucial role and it must be optimized in addition to its center, in contrast to the regular polygon.

Let us define $\mathbf{\Tilde{g}} = \begin{bmatrix}
    \Tilde{g}_1, \Tilde{g}_2, \Tilde{g}_3
\end{bmatrix}^T$ with $\Tilde{g}_3 = h_c$ as the location of line center. Thus, the $j$th element's location is given by
\begin{subequations}\label{eq:lineloc}
\begin{align}
    \label{eq:lineloc1} g_{j, 1} = \Tilde{g}_{1} - \big(\lfloor N/2\rfloor - j\big)\kappa\cos{\varphi},\\
    \label{eq:lineloc2} g_{j, 2} =  \Tilde{g}_{2} - \big(\lfloor N/2\rfloor - j\big)\kappa\sin{\varphi},
\end{align}
\end{subequations}
where $\varphi$ is the horizontal angle of the line. Then, we define $\Tilde{\mathbf{q}}_{j, i} = \mathbf{q}_i + \begin{bmatrix}
    \bigl(\bigl\lfloor{\frac{N}{2}\bigr\rfloor} - j\bigr)\kappa\cos{\varphi}, \bigl(\bigl\lfloor{\frac{N}{2}\bigr\rfloor} - j\bigr)\kappa\sin{\varphi}, 0
\end{bmatrix}^T$ and $\Tilde{d}_{j,i} = ||\Tilde{\mathbf{g}} - \Tilde{\mathbf{{q}}}_{j, i}||, \forall j,i$. Hereby, the problem becomes an SGP, which can be approximated using \eqref{prob_gp_ploy} and by replacing $\hat{q}_{j, i, u}$, $\hat{g}_u$, and $\hat{d}_{j,i}$ with $\Tilde{q}_{j, i, u}$, $\Tilde{g}_u$, and $\Tilde{d}_{j,i}$, respectively.

Herein, the optimization procedure is similar to Algorithm~\ref{sgp_poly_alg} and consists of iteratively approximating the problem until it converges to a local optimum. Notably, the solution obtained from \eqref{prob_gp_ploy} only specifies the center of the line for a given horizontal angle, thus, the best angle must still be found as it has a huge system performance impact. For this, we propose the heuristic represented in Algorithm~\ref{alg:lineshaped}, which consists of multiple search steps to find the local optimum solution for the location of the elements. Specifically,  $\varphi$ is increased iteratively and proportionally to $\zeta \ge 1$, and the location of the center and elements for the selected angle are obtained using \eqref{prob_gp_ploy} and \eqref{eq:lineloc}, respectively. Then, the line deployment leading to the best objective function value is selected.


\begin{algorithm}[t]
	\caption{Line-shaped radio stripe deployment.} \label{alg:lineshaped}
	\begin{algorithmic}[1]
            \State \textbf{Input:} $\Tilde{g}_{1}^{(0)}, {\Tilde{g}_{2}}^{(0)}$, $I_{max}$, $\epsilon$, $\zeta$ 
            \State \textbf{Output:} $\{\mathbf{g}^{\star}_{j}\}_{\forall j}$, $\{P_i\}_{\forall i}$
            \State \textbf{Initialize:} Compute $\{\Tilde{d}_{i}^{(0)}\}_{\forall i}$  for $\Tilde{g}_{1}^{(0)}, {\Tilde{g}_{2}}^{(0)}$, $iter = 1$, \\ $t =\infty $, $\Tilde{f} = 0$
            \For{$k = 1, \ldots, \zeta$}
                \State $\varphi \leftarrow \frac{k\pi}{\zeta} $
                \State Call Algorithm~\ref{sgp_poly_alg} with {Input:} $\Tilde{g}_{1}^{(0)}, {\Tilde{g}_{2}}^{(0)}$, $I_{max}$, $\epsilon$, \hspace{6mm} {Output:} $\{\mathbf{g}_{j}\}_{\forall j}$, $\{P_i\}_{\forall i}$ with \eqref{polyeqq} replaced by \eqref{eq:lineloc}
                \State Calculate $\min_i |\boldsymbol{\gamma}_i|^2$ using \eqref{eq:gain_simp} 
                \vspace{1mm}
                \If{$\min_i ||\boldsymbol{\gamma}_i||^2 \geq \Tilde{f}$}
                    \State $\Tilde{f} \leftarrow \min_i ||\boldsymbol{\gamma}_i||^2$, $\mathbf{g}^{\star}_{j} \leftarrow \mathbf{g}_{j}, \quad \forall j$
                \EndIf
            \EndFor
\end{algorithmic} 
\end{algorithm}

\vspace{-2mm}
\section{Numerical Analysis}\label{result}
\vspace{-1mm}
We consider a 64 m$^2$ indoor area with $h_c = 4$ m and 7 hotspots. We assume there is a single user per hotspot, which is randomly located within a 0.5-meter radius from the corresponding hotspot center in each Monte Carlo iteration out of 100. A fully-digital square array (referred to as Center-FD) and a square-shaped radio stripe (referred to as Center-Square), both located at the center of the area with $\lambda/2$ inter-element spacing, are used as benchmarks. Notice that the number of antenna elements (and RF~chains) in the fully-digital radio stripe network is determined by the antenna length and operation frequency, whereas the number of elements in the fully-digital square array is set to match its nearest square number. Furthermore, the performance indicator is the minimum power received by the users when exploiting both MRT-based and SDP-based precoders, which are obtained as in \cite{onellowcomp}. We adopt $\kappa = \lambda/2$ and set $b = 2$, $\zeta = 10$, $\omega = 1.1$, $I_{max} = 100$, $\epsilon = 10^{-6}$, and $\Tilde{P} = 1$ W.

\begin{figure}[t]
    \centering
    \vspace{-2mm}
    \includegraphics[width=0.8\columnwidth]{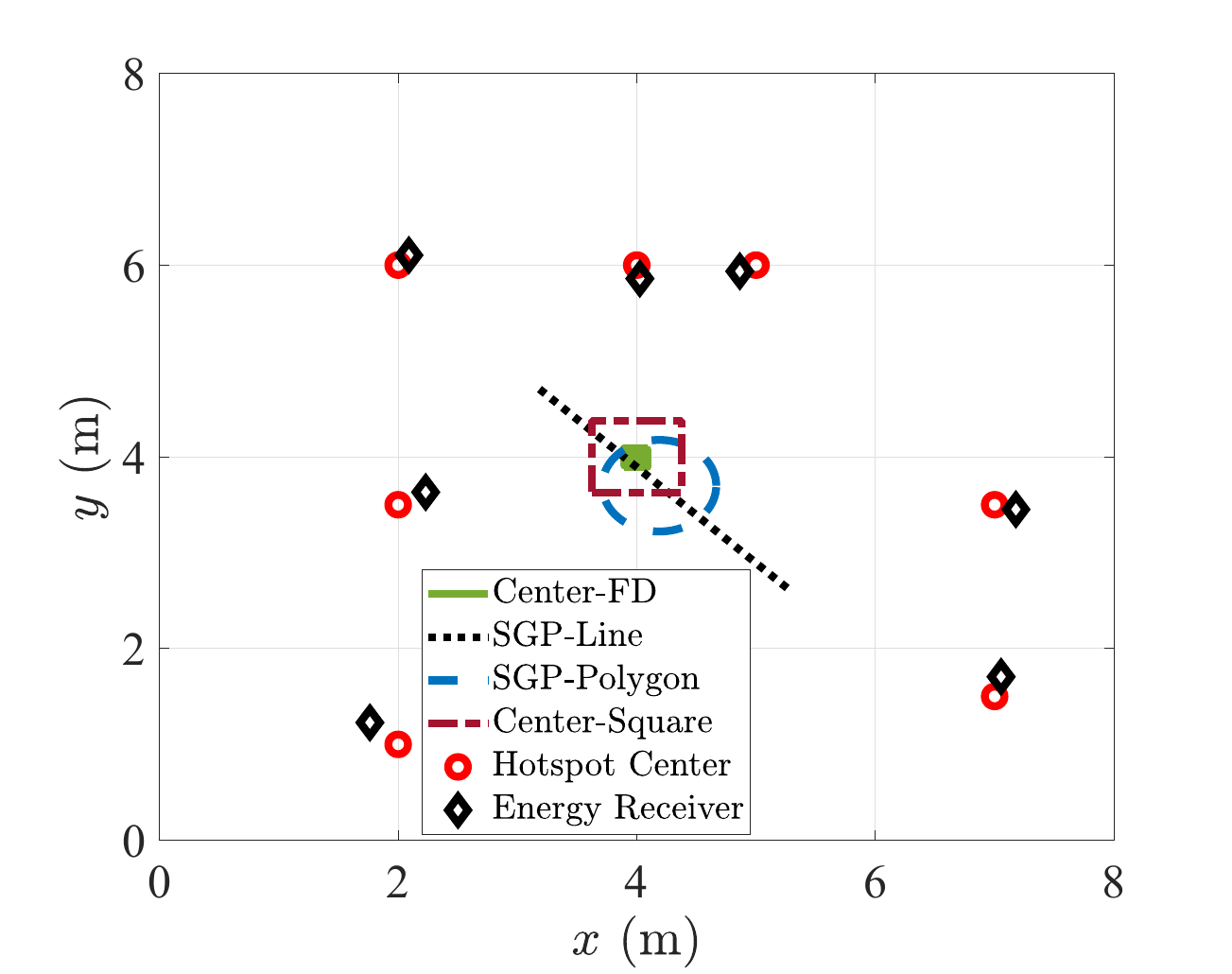}
    \vspace{-2mm}
    \caption{The 2D layout of the area illustrating the positions of both hotspots and devices, as well as the antenna deployments for $f = 10$ GHz and 3 m radio stripe length.}
    \label{fig:dist}
    \vspace{-5mm}
\end{figure}

Fig.~\ref{fig:dist} displays the 2D layout of the area, including the centers of the hotspots, a random realization of the device deployment, and the antenna deployments for 3 m cable length, $f = 10$ GHz, and $N = 200$. For better visualization, the figure only shows the shape on which the elements are uniformly positioned. Observe that both the polygon-shaped and line-shaped radio stripes are located in a way that can facilitate reaching all hotspots more efficiently.

\begin{figure}[t]
\centering
    \begin{subfigure}[b]{\columnwidth}
    \centering
        \includegraphics[width=0.86\columnwidth]{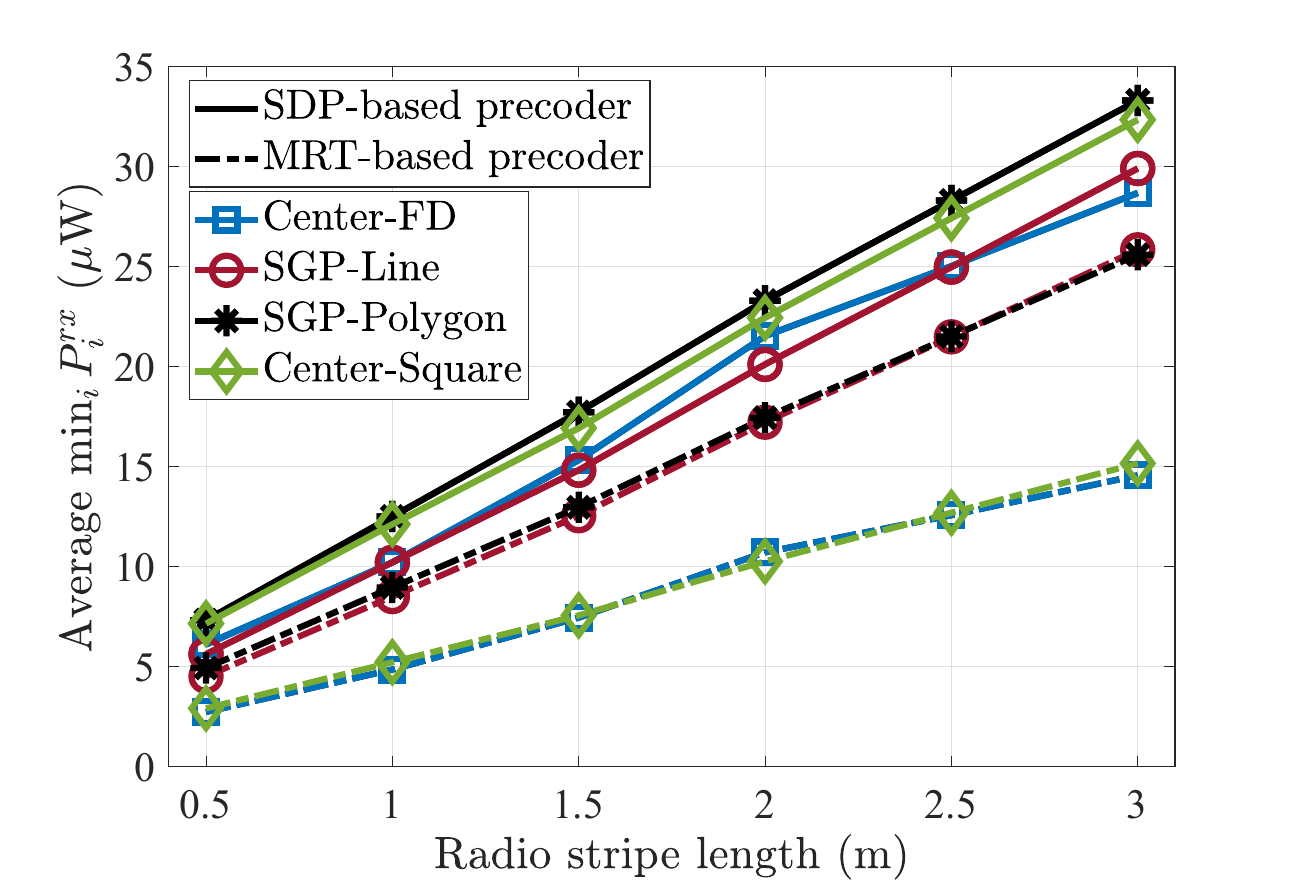}
    \end{subfigure}
    \vspace{-2mm}
    \begin{subfigure}[b]{\columnwidth}
    \centering
        \includegraphics[width=0.86\columnwidth]{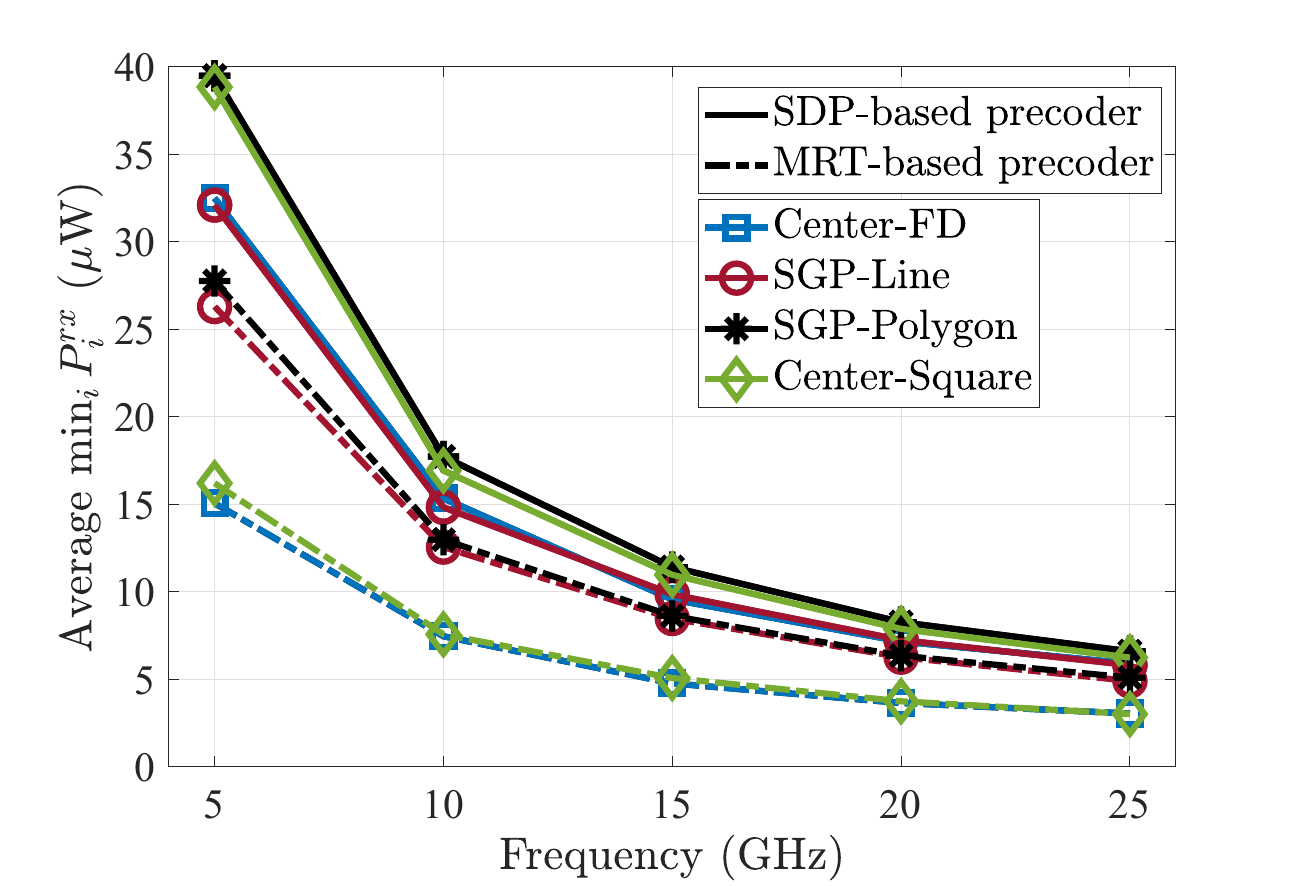}
    \end{subfigure}
    \vspace{-3mm}
    \caption{Average minimum power received by the users with MRT-based and SDP-based precoders as a function of (a) the radio stripe length with $f = 10$ GHz (top) and (b) the frequency for a 1.5 m radio stripe length (bottom).}
    \label{resultsfig}
    \vspace{-6mm}
\end{figure}

Fig.~\ref{resultsfig} illustrates the system performance as a function of $f$ and the radio stripe length. Although increasing the frequency provides more antenna elements for a given antenna length, it also increases the channel loss \cite{azarbahram2023energy}. Moreover, since the average distance between the users and the elements does not change much over frequency, the increased losses cause a reduction in minimum power received by the devices as shown in the results. As expected, the performance can be improved by utilizing a larger radio stripe length since a larger number of elements can be distributed in a wider area, and more degrees of freedom are provided for beamforming. Interestingly, the proposed deployments perform much better than the benchmarks for MRT-based precoders over both $f$ and radio stripe length, being the polygon-shaped deployment preferred over the line-shaped in this case. Meanwhile, the polygon-shaped radio stripe has the best performance also with SDP-based precoders. It is observed that Center-Square deployment performs similarly to the polygon-shaped deployment with SDP-based precoders, while line-shaped deployment degrades in this case, being similar to that of the Center-FD deployment. Notably, the performance of different radio stripe deployments is highly affected by the position of the hotspots. In this setup, the polygon-shaped and Center-Square deployments have a less effective distance to the users, thus, providing a better 3D beam focusing capability than the line-shaped deployment where some elements are far from specific users. Importantly, this study considered a relatively small setup, and in larger areas, the performance gap between radio stripes and Center-FD is expected to increase since the elements of Center-FD must be located close to each other, which leads to larger path losses due to the increased effective distance to the users. On the other hand, this issue can be mitigated in a radio stripe deployment by properly choosing the location and the shape.

\vspace{-2mm}
\section{Conclusion}\label{conclusion}
\vspace{-1mm}
In this paper, we considered a near-field RF-WPT system with a transmit radio stripe network and formulated its deployment problem aiming to maximize the minimum power received in the hotspots, i.e., the locations where the energy receivers are expected to be located. Furthermore, we solved the optimization problem for straight line-shaped and polygon-shaped radio stripes by relying on GP formulations. The numerical results evinced that the proposed radio stripe deployments can outperform a central fully-digital square array. Furthermore, it was observed that increasing the radio stripe length is beneficial while increasing the operational frequency leads to performance degradation due to larger channel losses. All in all, we showed the potential performance gains from deploying a properly shaped and located radio stripe network based on the characteristics of the deployment area and hotspot locations with respect to the conventional transmit architectures consisting of co-located antenna elements.

As a prospect for future research, one may investigate the optimization problem associated with free-form radio stripes without considering predefined shapes. Other interesting research directions could involve the integration of hotspots with different coverage areas and deploying multiple radio stripe networks for massive multi-antenna WPT systems.

\ifCLASSOPTIONcaptionsoff
  \newpage
\fi

\bibliography{ref_abbv}

\begin{thebibliography}{10}

\bibitem{intro1}
Z.~Zhang~\emph{et al.}, ``{6G wireless networks: vision, requirements, architecture, and key technologies},'' {\em IEEE Veh. Technol. Mag.}, vol.~14, no.~3, pp.~28--41, 2019.

\bibitem{intro2}
N.~H. Mahmood~\emph{et al.}, ``{Six key features of machine type communication in 6G},'' in {\em 2nd 6G SUMMIT}, pp.~1--5, 2020.

\bibitem{intro3}
O.~L.~A. López~\emph{et al.}, ``{Massive wireless energy transfer: enabling sustainable IoT toward 6G era},'' {\em IEEE Internet Things J.}, vol.~8, no.~11, pp.~8816--8835, 2021.

\bibitem{hybridbeamsurvey}
I.~Ahmed~\emph{et al.}, ``{A Survey on Hybrid Beamforming Techniques in 5G: Architecture and System Model Perspectives},'' {\em IEEE Commun. Surv. Tutor.}, vol.~20, no.~4, pp.~3060--3097, 2018.

\bibitem{IRS-basis}
Q.~Wu~\emph{et al.}, ``{Intelligent Reflecting Surface-Aided Wireless Communications: A Tutorial},'' {\em IEEE Trans Commun}, vol.~69, no.~5, pp.~3313--3351, 2021.

\bibitem{distributedWPT1}
C.~Zhang and G.~Zhao, ``{On the Deployment of Distributed Antennas of Power Beacon in Wireless Power Transfer},'' {\em IEEE Access}, vol.~6, pp.~7489--7502, 2018.

\bibitem{Osmel_Deployment}
O.~M. Rosabal~\emph{et al.}, ``{On the Optimal Deployment of Power Beacons for Massive Wireless Energy Transfer},'' {\em IEEE Internet Things J.}, vol.~8, no.~13, pp.~10531--10542, 2021.

\bibitem{intro_radiostripe}
G.~Interdonato~\emph{et al.}, ``{Ubiquitous cell-free Massive MIMO communications},'' {\em EURASIP J Wirel Commun Netw}, vol.~2019, p.~197, Aug 2019.

\bibitem{near-field}
H.~Zhang~\emph{et al.}, ``{Beam focusing for near-field multiuser MIMO communications},'' {\em IEEE Trans. Wirel. Commun.}, vol.~21, no.~9, pp.~7476--7490, 2022.

\bibitem{OnelRadioStripes}
O.~L.~A. López~\emph{et al.}, ``{Massive MIMO with radio stripes for indoor wireless energy transfer},'' {\em IEEE Trans. Wirel. Commun.}, vol.~21, no.~9, pp.~7088--7104, 2022.

\bibitem{shaik2021mmseRadiostripe}
Z.~H. Shaik~\emph{et al.}, ``{MMSE-optimal sequential processing for cell-free massive MIMO with radio stripes},'' {\em IEEE Trans Commun}, vol.~69, no.~11, pp.~7775--7789, 2021.

\bibitem{anetnna_radiation}
S.~W. Ellingson, ``{Path Loss in Reconfigurable Intelligent Surface-Enabled Channels},'' in {\em IEEE PIMRC}, pp.~829--835, 2021.

\bibitem{onellowcomp}
O.~L.~A. López~\emph{et al.}, ``{A Low-Complexity Beamforming Design for Multiuser Wireless Energy Transfer},'' {\em IEEE Wireless Commun. Lett.}, vol.~10, no.~1, pp.~58--62, 2021.

\bibitem{gp_boyd}
S.~Boyd~\emph{et al.}, ``{A tutorial on geometric programming},'' {\em Optimization and Engineering}, vol.~8, pp.~67--127, Mar 2007.

\bibitem{cvxref}
M.~Grant and S.~Boyd, ``{CVX}: Matlab software for disciplined convex programming, version 2.1.'' \url{http://cvxr.com/cvx}, Mar. 2014.

\bibitem{boyd2004convex}
S.~P. Boyd and L.~Vandenberghe, {\em {Convex optimization}}.
\newblock Cambridge university press, 2004.

\bibitem{azarbahram2023energy}
A.~Azarbahram~\emph{et al.}, ``{Energy Beamforming for RF Wireless Power Transfer with Dynamic Metasurface Antennas},'' 2023.

\end{thebibliography}
\bibliographystyle{ieeetr}

\end{document}